\documentstyle[12pt,a4wide]{article}

\newcommand{\nin}{\noindent}
\newcommand{\be}{\begin{equation}}
\newcommand{\ee}{\end{equation}}
\newcommand{\bea}{\begin{eqnarray}}
\newcommand{\eea}{\end{eqnarray}}

\newcommand{\nonu}{\nonumber\\}

\begin{document}

\begin{titlepage}

\begin{flushright}
hep-th/0207232
\end{flushright}

\begin{center}
\textbf{\Large Induced $N=2$ composite supersymmetry\\
in 2+1 dimensions}

\vspace{1cm}

{\bf J.~Alexandre}, {\bf N.E.~Mavromatos} and {\bf Sarben Sarkar}\\
Department of Physics, Theoretical Physics Group, King's College London,\\
Strand, London WC2R 2LS, UK

\vspace{2cm}

\textbf{Abstract}
\end{center}

\noindent Starting from $N=1$ scalar supermultiplets in 2+1 dimensions, we
build explicitly the composite superpartners which define a $N=2$
superalgebra induced by the initial $N=1$ supersymmetry. The occurrence of
this extension is linked to the topologically conserved current out of which
the composite superpartners are constructed.

\end{titlepage}

\bigskip The study of $2+1$ dimensional gauge theories has been motivated by
the search for an understanding of confinement \cite{conf}. The degree to
which they can be of help in understanding confinement at zero temperature
in three space dimensions is unclear because the nature and possible
configurations of textures, instantons and other non-perturbative features
such as the locality of disorder variables are very different. Their
attraction is that their comparative simplicity allows a much more complete
understanding of the confinement phenomenon. However another reason for
studying such theories is the interest in planar system in high temperature
superconductors \cite{var} where the phases are believed not to be of
conventional type . In particular string-like structures which are features
of confinement in the gauge theories occur as inhomogeneities in the charge
and magnetic order \cite{stripe} in certain parameter regimes of these
materials. Both in the study of confinement and the condensed matter
analogues, some of the reason for controversy is the room for error in the
calculations that can be performed due to the intrinsic limitations of the
methods themselves. It is thus important to pursue models with exact
solutions. Unlike for $1+1$ dimensions where infinite-dimensional groups and
the factorizability of the S-matrix are at play \cite{Zam},\ dynamical
supersymmetry is an important ingredient for obtaining exact information
about the phases of the theory \cite{susy}. Although $N=1$ symmetry has been
demonstrated for a condensed matter model \cite{mav}, a lack of
holomorphicity properties has meant that exact non-perturbative information
is not available. The situation is much improved for $N=2$ supersymmetry.
This paper demonstrates how that $N=1$ supersymmetry, which was in terms of
constituent fields, can be elevated to a $N=2$ supersymmetry in terms of
suitably constructed composite fields.

Our results may be of relevance in attempts towards an analytic
understanding at a non-perturbative level of the dynamics of
strongly-correlated electron systems, with relevance to high-temperature
superconductivity and more generally to antiferromagnetism. Also our model
may be viewed as a toy model for an understanding of ideas related to the
so-called \textit{scaleless limit} of gauge theories~\cite{bjorken}, where
the gauge fields appear dynamically from more fundamental interacting
constituents in the theory.

However, in the four-dimensional setting of \cite{bjorken}, the emergent
gauge bosons (photons) appeared as Goldstone bosons of a spontaneous
breakdown of Lorentz symmetry, associated with non-zero vacuum expectation
values (v.e.v.) of vector fields linearising the four-fermion Thirring
interactions of the model. In three space-time dimensions, on the other
hand, the photons have only a single degree of freedom and hence they are
allowed to get a v.e.v without breaking Lorentz symmetry. It is this fact
that allows the extension of such ideas in (2+1)-dimensions to incorporate
supersymmetry, which is intimately related to Lorentz invariance. From a
physical point of view, with relevance to strongly-correlated electrons, we
have Lorentz symmetry since we restrict our attention to excitations near
nodes of the fermi surface of such systems~\cite{mav}; the r\^{o}le of the
velocity of `light' is played by the fermi velocity at the node.

We will first summarize the results of \cite{compos}, dealing with distinct $%
N=1$ quadratic composite supermultiplets and then describe the procedure of
adding higher order composites so as to generate the coupling between the
scalar and vector supermultiplets. This coupling implies the $N=2$ structure
of the transformations. Eventually, we give the explicit construction of the
quartic composites and show that their existence is intimately linked to the
topologically conserved current.

\bigskip

\noindent We use a representation of the Clifford algebra where all the $%
\gamma$-matrices are real: $\gamma^0=-i\sigma^2$, $\gamma^1=\sigma^1$ and $%
\gamma^2=\sigma^3$ and they satisfy

\begin{eqnarray}  \label{propgamma}
\left\{\gamma^\mu,\gamma^\nu\right\}&=&2g^{\mu\nu},~~~~\mbox{with}%
~~g^{\mu\nu}=(-1,1,1)  \nonumber \\
\left[\gamma^\mu,\gamma^\nu\right]&=&-2\epsilon^{\mu\nu\rho}\gamma_\rho.
\end{eqnarray}

\noindent We start from two complex $N=1$ scalar supermultiplets $%
(z_a,\psi_a,f_a)$, $a=1,2$, containing the constituent fields which
transform as

\begin{eqnarray}  \label{susytransfoinit}
\delta z_a&=&\overline\varepsilon\psi_a  \nonumber \\
\delta\psi_a&=&\hskip .25cm/\hskip -.25cm\partial
z_a\varepsilon+f_a\varepsilon  \nonumber \\
\delta f_a&=&\overline\varepsilon\hskip .25cm/\hskip -.25cm\partial\psi_a,
\end{eqnarray}

\noindent where $\varepsilon $ is a real Grassmann parameter and $\overline{%
\varepsilon }=\varepsilon ^{T}\gamma ^{0}$. We will use the following
essential properties concerning the scalar quantities:

\begin{eqnarray}  \label{scalprop}
\overline\psi_1\psi_2&=&(\overline\psi_2\psi_1)^\star  \nonumber \\
\overline\psi_1\gamma_\mu\psi_2&=&-(\overline\psi_2\gamma_\mu\psi_1)^\star,
\end{eqnarray}

\noindent and the $2\times 2$ matrix \cite{compos}:

\begin{equation}  \label{property}
\psi_2\overline\psi_1=-\frac{1}{2}\left(\overline\psi_1\psi_2+(\overline
\psi_1\gamma_\mu\psi_2)\gamma^\mu\right),
\end{equation}

\noindent where $\overline{\psi}_a=\psi_a^\dagger\gamma ^{0}$.

It was found in \cite{compos} that the quadratic composites

\begin{eqnarray}  \label{defcomp}
\phi&=&\overline\psi_1\psi_2 \\
A_\mu&=&\overline\psi_1\gamma_\mu\psi_2-z_1^\star\partial_\mu
z_2+z_2\partial_\mu z_1^\star  \nonumber
\end{eqnarray}

\noindent belong respectively to a $N=1$ scalar supermultiplet $(\phi ,\psi
,f)$ and a $N=1$ vector supermultiplet $(A_{\mu },\chi )$ where, in terms of
the constituent fields, the composite superpartners read as follows:

\begin{eqnarray}  \label{quadpartners}
\psi&=&(f_1^\star-\hskip .25cm/\hskip -.25cm\partial z_1^\star)\psi_2+(f_2- %
\hskip .25cm/\hskip -.25cm\partial z_2)\psi_1^\star  \nonumber \\
\chi&=&(f_1^\star+\hskip .25cm/\hskip -.25cm\partial z_1^\star)\psi_2-(f_2+ %
\hskip .25cm/\hskip -.25cm\partial z_2)\psi_1^\star  \nonumber \\
f&=&2(f_1^\star f_2-\partial_\mu z_1^\star\partial^\mu z_2)-\overline\psi_1 %
\hskip .25cm/\hskip -.25cm\partial\psi_2 -(\overline\psi_2\hskip .25cm/ %
\hskip -.25cm\partial\psi_1)^\star.
\end{eqnarray}

\noindent The transformation of $A_{\mu }$ has actually the expected form up
to a gauge transformation, which implies that it must be a gauge field. The
fields defined in Eqs.(\ref{defcomp}) are complex; although they are not the
physical degrees of freedom, they have the generic composite structure that
we want to study. The physical degrees of freedom form $SU(2)$ triplets of
composites which are given in \cite{compos}, such that we have three $N=1$
scalar supermultiplets and three $N=1$ vector supermultiplets. From the
three scalar composite supermultiplets, we form one real and one complex $%
N=1 $ scalar supermultiplets, that we denote by $(\phi _{r},\psi _{r},f_{r})$
and $(\phi _{c},\psi _{c},f_{c})$ respectively. The $N=1$ composite
supersymmetric transformations, found in \cite{compos}, are then

\begin{eqnarray}  \label{transfocompN1}
\delta\phi_r&=&\overline\varepsilon\psi_r,~~~~\delta\psi_r=(\hskip .25cm/ %
\hskip -.25cm\partial\phi_r+f_r)\varepsilon, ~~~~\delta
f_r=\overline\varepsilon\hskip .25cm/\hskip -.25cm\partial\psi_r,  \nonumber
\\
\delta\phi_c&=&\overline\varepsilon\psi_c,~~~~\delta\psi_c=(\hskip .25cm/ %
\hskip -.25cm\partial\phi_c+f_c)\varepsilon, ~~~~\delta
f_c=\overline\varepsilon\hskip .25cm/\hskip -.25cm\partial\psi_c,  \nonumber
\\
\delta A_\mu^a&=&\overline\varepsilon\gamma_\mu\chi^a,~~~~\delta\chi^a= -%
\frac{1}{2}F_{\mu\nu}^a\gamma^\mu\gamma^\nu\varepsilon,
\end{eqnarray}

\noindent where $F_{\mu\nu}^a$ is the \textit{Abelian} field strength of $%
A_\mu^a$ (quadratic in the constituent fields), for each gauge index $%
a=1,2,3 $ and the gauginos $\chi^a$ are real.

\bigskip

In order to elevate the $N=1$ supersymmetry to a $N=2$ (Abelian)
supersymmetry, we should couple the complex composite scalar supermultiplet
to one of the composite vector supermultiplets, which we denote as $%
(A_\mu,\chi)$. For this, we will construct explicitly the covariant
derivative of $\phi_c$ by adding to the complex quadratic scalar field a
\textit{higher order composite} which will generate the minimal coupling. We
take into account the quartic contribution only and neglect the higher
orders: we are looking for a quartic composite scalar $M$ whose
supersymmetric transformations generate the quartic fermion $\Lambda$ and
the quartic auxiliary field ${\cal F}$ such that, under the
transformations (\ref{susytransfoinit}), one has

\begin{eqnarray}  \label{quartictransfo}
\delta M&=&\overline\varepsilon\Lambda  \nonumber \\
\delta\Lambda&=&\left(-i\hskip .25cm/\hskip -.25cm A\phi_c+\hskip .25cm/ %
\hskip -.25cm\partial M+{\cal F}\right)\varepsilon  \nonumber \\
\delta{\cal F}&=&\overline\varepsilon(-i\hskip .25cm/\hskip -.25cm
A\psi_c+\hskip .25cm/\hskip -.25cm\partial\Lambda)
\end{eqnarray}

\noindent We shall find that the transformation (\ref{quartictransfo}) of
the fermion is satisfied in the gauge defined by the \textit{complex}
equation

\begin{equation}  \label{gaugecond}
\partial^\nu\left(A_\nu\Box\phi_c\right)=0,
\end{equation}

\noindent which consequently implies two gauge conditions. This is possible
in (2+1) dimensions, since a gauge field has one physical degree of freedom %
\cite{binegar}. The transformation (\ref{quartictransfo}) of the auxiliary
field will be satisfied up to irrelevant operators which are higher order
derivative operators. We will neglect these derivative interactions between
the composites. This approximation is in the spirit of the low-energy
effective theories that concern us. Such terms will not affect the infrared
universality class of the model in which we are interested.

Making the substitutions:

\begin{eqnarray}  \label{substitut}
\phi_c&\to&\Phi=\phi_c+gM  \nonumber \\
\psi_c&\to&\Psi=\psi_c+g\Lambda  \nonumber \\
f_c&\to&F=f_c+g{\cal F},
\end{eqnarray}

\noindent where $g$ is a dimensionful constant, we shall then obtain the
expected covariant derivatives

\begin{eqnarray}  \label{covderiv}
D_\mu\Phi&=&(\partial_\mu-igA_\mu)\Phi  \nonumber \\
&=&\partial_\mu\phi_c-igA_\mu\phi_c +g\partial_\mu M +%
\mbox{sixth order
composite}  \nonumber \\
D_\mu\Psi&=&(\partial_\mu-igA_\mu)\Psi  \nonumber \\
&=&\partial_\mu\psi_c-igA_\mu\psi_c +g\partial_\mu \Lambda +
\mbox{sixth
order composite}
\end{eqnarray}

\noindent in the (supersymmetry) transformation laws of the complex fermion
$\Psi$ and auxiliary field $F$ respectively. Next, we shall construct the
complex gaugino $\lambda=\psi_r+i\chi$ and obtain from Eqs.(\ref
{transfocompN1}) the following transformations

\begin{eqnarray}\label{transfofinal}
\delta \Phi &=&\overline{\varepsilon }\Psi ,~~~~~~~~\delta \Psi =(\hskip
.25cm/\hskip-.25cmD\Phi +F)\varepsilon ,  \nonu
\delta \phi _{r} &=&\overline{\varepsilon }\psi _{r},~~~~~~~~\delta A_{\mu }=
\overline{\varepsilon }\gamma _{\mu }\chi ,  \nonu
\delta \lambda &=&\left( \hskip.25cm/\hskip-.25cm\partial \phi _{r}+f_{r}-
\frac{i}{2}F_{\mu \nu }\gamma ^{\mu }\gamma ^{\nu }\right) \varepsilon\nonu
\delta f_{r} &=&\overline{\varepsilon }\hskip.25cm/\hskip-.25cm\partial \psi
_{r}~~~~~~~~\delta F\simeq \overline{\varepsilon }\hskip.25cm/\hskip
-.25cmD\Psi ,
\end{eqnarray}

\nin where the reason for the symbol $\simeq$
in the transformation of the auxiliary field $F$ will
become clear later on (see discussion after Eq.(\ref{defS})).

The above equations (\ref{transfofinal}) constitute an off-shell
set of $N=2$ transformations if and only if
the parameter $\varepsilon$ is complex,
i.e. when the transformations (\ref{transfofinal}) are
followed by a phase rotation of
the spinors $\lambda$ and $\Psi$ (note that
a phase rotation of $\lambda$ is
equivalent to the same phase rotation for $\chi$ and $\psi_r$, which
is necessary in order
to keep the fields $\phi_r$ and $A_\mu$ real).
In \cite{edelstein}, where ``elementary''
and not composite fields are considered,
the authors write a set of {\it on-shell} $N=2$ transformations in the
context of the Abelian Higgs model and use the equations of motion of the
auxiliary fields. These auxiliary fields
are then expressed in terms of the scalar fields $\phi_r$ and $\Phi$
and the coupling $g'$ of the Higgs potential~\cite{edelstein}.
This Higgs self-coupling $g'$
{\it does not} appear in our purely algebraic considerations
above. Nevertheless, at a Lagrangian level,
the $N=2$ superalgebra structure imposes a
relation~\cite{edelstein}
between the couplings $g$ and $g'$.
Since we are dealing here with only algebraic aspects of the composite
supersymmetry, we shall not go into such detailed discussions on the
dynamics of the model
and the associated
physical consequences for
strongly-correlated electron systems. Such issues have been briefly
mentioned in \cite{compos}, and a more detailed analysis
will be postponed to a forthcoming publication.

We also remark that the extension of the composite $N=1$ transformations
(\ref{transfocompN1}) to a $N=2$ \textit{non-Abelian} superalgebra would
involve too many gauge conditions. Besides the addition of a quartic scalar,
this extension requires quartic gauge fields ${\cal A}_\mu^a$  in
order to generate the covariant derivatives.The ${\cal A}_\mu^a$
would transform into quartic gauginos $\Xi^a$ as follows:

\begin{eqnarray}  \label{nonAbelian}
\delta{\cal A}_\mu^a&=&\overline\varepsilon\gamma_\mu\Xi^a  \nonumber \\
\delta\Xi^a&=&f^{abc}\hskip .25cm/\hskip -.25cm A^b\hskip .25cm/\hskip
-.25cm A^c\varepsilon,
\end{eqnarray}

\noindent so as to generate the non-Abelian field strength in the
transformation of the gauginos ($f^{abc}$ are the $SU(2)$ structure
constants). We found that the transformations (\ref{nonAbelian}) of the
quartic gauginos are satisfied provided we impose two new constraints on
each colour of the gauge field. This shows that a non-Abelian model of the
type considered in \cite{affleck}, cannot be obtained with the composite
procedure developed here.

Finally, we note that no new constraints arise if we consider higher order
composites: the quartic composite can be seen as the truncation of a series
of composite operators which lead to the exact covariant derivatives if they
are ressummed. The gauge condition (\ref{gaugecond}) gives rise to a gauge
fixing term in the Lagrangian which is also the truncation of a series of
gauge fixing terms.

\bigskip

We now proceed to construct explicitly the quartic-order superpartners which
lead to the transformations (\ref{transfofinal}). First we note that, in
order for the expression (\ref{covderiv}) of the covariant derivative to be
consistent, the mass dimension of $g$ should be -1, such that the mass
dimension of $M$ should be 3. Let us consider the following possibilities

\begin{eqnarray}  \label{possM}
\Box M^{(1)}&=&\phi_c\partial^\mu A_\mu  \nonumber \\
\Box M^{(2)}&=&A_\mu\partial^\mu\phi_c  \nonumber \\
\Box M^{(3)}&=&\overline\psi_c^\star\chi.
\end{eqnarray}

\noindent The supersymmetric transformation of $M^{(1)}$ under (\ref%
{susytransfoinit}) leads to $\delta
M^{(1)}=\overline\varepsilon\Lambda^{(1)} $ with

\begin{equation}
\Box\Lambda^{(1)}=\partial^\mu A_\mu\psi_c+\phi_c\hskip .25cm/\hskip %
-.25cm\partial\chi.
\end{equation}

\noindent Taking into account the properties (\ref{scalprop}), we find that
the transformation of $\Box\Lambda^{(1)}$ reads

\begin{eqnarray}
\delta\Box\Lambda^{(1)}&=& \left(\hskip .25cm/\hskip -.25cm\partial\left(
\phi_c\partial^\mu A_\mu\right) -\phi_c\hskip .25cm/\hskip -.25cm\partial %
\hskip .25cm/\hskip -.25cm\partial\hskip .25cm/\hskip -.25cm
A+f_c\partial^\mu A_\mu\right)\varepsilon  \nonumber \\
&&+\left(\hskip .25cm/\hskip -.25cm\partial\chi\overline\psi_c^\star
-\psi_c\partial^\nu\overline\chi\gamma_\nu\right)\varepsilon.
\end{eqnarray}

\noindent Finally, the property (\ref{property}) gives

\begin{equation}
\delta\Box\Lambda^{(1)}=\hskip .25cm/\hskip -.25cm\partial\Box
M^{(1)}\varepsilon +\left(f_c\partial^\mu A_\mu-\phi_c\Box \hskip .25cm/ %
\hskip -.25cm A\right)\varepsilon -\left(\overline\psi_c^\star\hskip .25cm/ %
\hskip -.25cm\partial\chi\right)\varepsilon.
\end{equation}

\noindent If we define $\delta M^{(2)}=\overline\varepsilon\Lambda^{(2)}$
and $\delta M^{(3)}=\overline\varepsilon\Lambda^{(3)}$ and proceed in a
similar way, we find

\begin{eqnarray}
\Box\Lambda^{(2)}&=&\hskip .25cm/\hskip -.25cm\partial\phi_c\chi+A_\mu
\partial^\mu\psi_c  \nonumber \\
\Box\Lambda^{(3)}&=&(f_c-\hskip .25cm/\hskip -.25cm\partial\phi_c)\chi + (%
\hskip .25cm/\hskip -.25cm\partial\hskip .25cm/\hskip -.25cm A-\partial_\mu
A^\mu)\psi_c,
\end{eqnarray}

\noindent and the transformations of $\Box\Lambda^{(2)}$ and $%
\Box\Lambda^{(3)}$ are

\begin{eqnarray}
&~& \delta\Box\Lambda^{(2)}=\hskip .25cm/\hskip -.25cm\partial\Box
M^{(2)}\varepsilon
-\left(\partial^\mu\overline\psi_c^\star\gamma_\mu\chi\right)\varepsilon \\
&~&~~+\left(A_\mu\partial^\mu f_c-\partial^\mu\phi_c\hskip .25cm/\hskip %
-.25cm\partial A_\mu +\hskip .25cm/\hskip -.25cm\partial\phi_c\partial^\mu
A_\mu-\hskip .25cm/\hskip -.25cm\partial\phi_c\hskip .25cm/\hskip %
-.25cm\partial\hskip .25cm/\hskip -.25cm A\right)\varepsilon  \nonumber \\
&~& \delta\Box\Lambda^{(3)}=\hskip .25cm/\hskip -.25cm\partial\Box
M^{(3)}\varepsilon
+\partial^\mu\left(\overline\psi_c^\star\gamma_\mu\chi\right)\varepsilon
\nonumber \\
&~&~~+\left(2\hskip .25cm/\hskip -.25cm\partial\phi_c\hskip .25cm/\hskip %
-.25cm\partial\hskip .25cm/\hskip -.25cm A+2\hskip .25cm/\hskip %
-.25cm\partial A_\mu\partial^\mu\phi_c -2\partial_\mu\phi_c\partial^\mu %
\hskip .25cm/\hskip -.25cm A-2\hskip .25cm/\hskip -.25cm\partial\phi_c
\partial_\mu A^\mu\right)\varepsilon .  \nonumber
\end{eqnarray}

\noindent If we define the scalar $\tilde M$ to be the linear combination

\begin{equation}
\tilde M=M^{(1)}+2M^{(2)}+M^{(3)},
\end{equation}

\noindent and note that $\delta\tilde M=\overline\varepsilon\tilde\Lambda=
\overline\varepsilon(\Lambda_1+2\Lambda_2+\Lambda_3)$, then we find

\begin{equation}  \label{gaga}
\delta\Box\tilde\Lambda=\left(-\phi_c\Box\hskip .25cm/\hskip -.25cm A
-2\partial_\mu\phi_c\partial^\mu\hskip .25cm/\hskip -.25cm A+\hskip .25cm/ %
\hskip -.25cm\partial\Box\tilde M +\Box\tilde{\cal F}\right)\varepsilon
\nonumber \\
\end{equation}

\noindent where the auxiliary field $\tilde{\cal F}$ satisfies

\begin{equation}  \label{auxtilde}
\Box\tilde{\cal F}=2A_\mu\partial^\mu f_c+f_c\partial^\mu A_\mu
-\partial^\mu\overline\psi_c^\star\gamma_\mu\chi.
\end{equation}

\noindent In order to generate the expected minimal coupling
$\Box\left(\phi_c\hskip .25cm/ \hskip -.25cm A\right)$, we still need to add
to (\ref{gaga}) the operator $-\Box\phi_c \hskip .25cm/\hskip -.25cm A$,
which will be obtained by the introduction of the scalar $M^{(4)} $ defined
by the following equation

\begin{equation}
\partial^\rho\Box M^{(4)}=\epsilon^{\mu\nu\rho}f_c\partial_\mu A_\nu.
\end{equation}

\noindent Note that the occurrence of this scalar is specific to 2+1
dimensions, as it is proportional to the topologically conserved current
$J^\rho=\epsilon^{\mu\nu\rho}\partial_\mu A_\nu$, which plays a central role
in the elevation of an $N=1$ supersymmetry to an extended $N=2$
supersymmetry \cite{hlousek}. If we consider the identity (\ref{propgamma}),
we find that

\begin{equation}
\hskip .25cm/\hskip -.25cm\partial\hskip .25cm/\hskip -.25cm\partial \Box
M^{(4)}=\Box^2 M^{(4)}= \hskip .25cm/\hskip -.25cm\partial\left(f_c(
\partial^\nu A_\nu-\hskip .25cm/\hskip -.25cm\partial\hskip .25cm/\hskip %
-.25cm A)\right),
\end{equation}

\noindent which will be used in the supersymmetric transformations of $%
M^{(4)} $. The gaugino $\Lambda^{(4)}$, defined by $\delta
M^{(4)}=\overline\varepsilon\Lambda^{(4)}$, satisfies

\begin{equation}
\partial^\rho\Box\Lambda^{(4)}=\epsilon^{\mu\nu\rho} \left(\partial_\mu
A_\nu \hskip .25cm/\hskip -.25cm\partial\psi_c+f_c\gamma_\nu\partial_\mu\chi
\right),
\end{equation}

\noindent such that

\begin{equation}
\hskip .25cm/\hskip -.25cm\partial\Box\Lambda^{(4)}=(\partial^\nu A_\nu- %
\hskip .25cm/\hskip -.25cm\partial\hskip .25cm/\hskip -.25cm A)\hskip .25cm/ %
\hskip -.25cm\partial \psi_c-2f_c\hskip .25cm/\hskip -.25cm\partial\chi.
\end{equation}

\noindent The supersymmetric transformation of $\Lambda^{(4)}$ then gives,
when using (\ref{property}),

\begin{eqnarray}
&&\delta\left(\hskip .25cm/\hskip -.25cm\partial\Box\Lambda^{(4)}\right) \\
&&=\left((\partial^\nu A_\nu-\hskip .25cm/\hskip -.25cm\partial\hskip .25cm/
\hskip -.25cm A)(\hskip .25cm/\hskip -.25cm\partial f_c+\Box\phi_c) -2f_c
\hskip .25cm/\hskip -.25cm\partial(\partial^\nu A_\nu-\hskip .25cm/\hskip
-.25cm\partial\hskip .25cm/\hskip -.25cm A)\right)\varepsilon  \nonumber \\
&&~~+\gamma_\rho\left(\partial_\sigma\overline\psi_c^\star\gamma^\rho
\partial^ \sigma\chi -\partial^\rho\overline\psi_c^\star\hskip .25cm/\hskip
-.25cm\partial\chi\right)\varepsilon
+\epsilon^{\mu\nu\rho}\left(\partial_\nu\overline\psi_c^\star\partial_\mu
\chi\right) \gamma_\rho\varepsilon  \nonumber \\
&&=\hskip .25cm/\hskip -.25cm\partial\left(-\Box\phi_c\hskip .25cm/\hskip
-.25cm A+\hskip .25cm/\hskip -.25cm\partial\Box M^{(4)} +\Box{\cal F}
^{(4)} \right)\varepsilon
+\partial^\nu\left(A_\nu\Box\phi_c\right)\varepsilon ,\nonumber
\end{eqnarray}

\noindent where the quartic auxiliary field ${\cal F}^{(4)}$ satisfies

\begin{eqnarray}  \label{aux4}
\partial^\rho\Box{\cal F}^{(4)}&=&\epsilon^{\mu\nu\rho}A_\mu\partial_\nu
\Box\phi_c +2\partial_\nu\left(f_c\partial^{[\nu} A^{\rho]}\right)
+f_c\left(\Box A^\rho-\partial^\rho\partial^\nu A_\nu\right)  \nonumber \\
&&+\partial_\sigma\overline\psi_c^\star\gamma^\rho\partial^\sigma\chi
-\partial^\rho\overline\psi_c^\star\hskip .25cm/\hskip -.25cm\partial\chi
+\epsilon^{\mu\nu\rho}\partial_\nu\overline\psi_c^\star\partial_\mu\chi.
\end{eqnarray}

\noindent We now assume the gauge choice

\begin{equation}  \label{gauge}
\partial^\nu\left(A_\nu\Box\phi_c\right)=0,
\end{equation}

\noindent which allows us to write the transformation of $\Box\Lambda^{(4)}$
in the expected form, i.e.

\begin{equation}
\delta\left(\Box\Lambda^{(4)}\right)= \left(-\Box\phi_c\hskip .25cm/\hskip
-.25cm A+\hskip .25cm/\hskip -.25cm\partial\Box M^{(4)}+\Box{\cal F}
^{(4)}\right)\varepsilon,
\end{equation}

\noindent such that the final quartic spinor that has the expected
transformation (\ref{quartictransfo}) is

\begin{equation}
\Lambda=i(\tilde\Lambda+\Lambda^{(4)}),
\end{equation}

\noindent and the quartic scalar $M$ and auxiliary field ${\cal F}$ are
given by

\begin{eqnarray}
M&=&i(\tilde M+M^{(4)})  \nonumber \\
{\cal F}&=&i(\tilde{\cal F}+{\cal F}^{(4)}).
\end{eqnarray}

\noindent Let us now consider the auxiliary field ${\cal F}$. It can be
easily checked that

\begin{eqnarray}
-i\Box^2{\cal F}&=&\Box\left(2A_\mu\partial^\mu f_c+f_c\partial^\mu
A_\mu\right) +\epsilon^{\mu\nu\rho}\partial_\rho A_\mu\partial_\nu\Box\phi_c
\nonumber \\
&&-\partial^\mu\left(\Box\overline\psi_c^\star\gamma_\mu\chi
+\partial^\nu\overline\psi_c^\star\gamma_\nu\partial_\mu\chi\right).
\end{eqnarray}

\noindent The supersymmetric transformation of $\Box^2{\cal F}$ is then

\begin{equation}
\delta\left(\Box^2{\cal F}\right)=\overline\varepsilon\Box^2(-i\hskip
.25cm/\hskip -.25cm A\psi_c+\hskip .25cm/\hskip -.25cm\partial\Lambda)
+\overline\varepsilon S,
\end{equation}

\noindent where the spinor $S$ reads

\begin{eqnarray}\label{defS}
S&=&\partial^\mu f_c\partial_\mu\hskip .25cm/\hskip -.25cm\partial\chi-
\hskip .25cm/\hskip -.25cm\partial f_c\Box\chi
-\Box(\phi_c\Box\chi+2\partial^\mu\phi_c\partial_\mu\chi)  \nonumber \\
&&+\Box\left(2\Box \hskip .25cm/\hskip -.25cm A\psi_c+4\partial^\mu\hskip
.25cm/\hskip -.25cm A\partial_\mu\psi_c+\hskip .25cm/\hskip -.25cm
A\Box\psi_c -\hskip .25cm/\hskip -.25cm\partial
A^\mu\partial_\mu\psi_c\right)  \nonumber \\
&&+\partial^\sigma(\partial^\nu A_\nu-\hskip .25cm/\hskip -.25cm\partial
\hskip .25cm/\hskip -.25cm A)(\gamma_\sigma\Box\psi_c -\partial_\sigma\hskip
.25cm/\hskip -.25cm\partial\psi_c)  \nonumber \\
&&-\epsilon^{\mu\nu\rho}\hskip .25cm/\hskip -.25cm\partial\partial_\mu
A_\nu\partial_\rho\hskip .25cm/\hskip -.25cm\partial\psi_c
\end{eqnarray}

\noindent As we can now see, setting $S$ to zero amounts to imposing irrelevant
constraints on the dynamics and
neglecting higher order derivative operators, which can be omitted in the
context of low energy effective theories,
given that they do not affect the infrared
universality class of the model. Hence, up to such terms, one
obtains the following supersymmetry transformation
for ${\cal F}$:

\begin{equation}
\delta{\cal F}\simeq\overline\varepsilon(-i\hskip .25cm/\hskip -.25cm
A\psi_c+\hskip .25cm/\hskip -.25cm\partial\Lambda).
\end{equation}

\nin which, on account of (\ref{substitut}) and (\ref{transfocompN1}),
yields
the standard $N=2$ supersymmetry
transformation (\ref{transfofinal}) for the auxiliary
field $F$.

Finally, the quartic scalar $M$ that we were looking for satisfies:

\begin{equation}
-i\partial^\rho\Box M=\epsilon^{\mu\nu\rho}f_c\partial_\mu A_\nu
+\partial^\rho\left(2\partial^\mu\phi_c A_\mu+\phi_c\partial^\mu
A_\mu+\overline\psi^\star_c\chi\right).
\end{equation}
This completes the off-shell closure of the $N=2$ supersymmetric algebra for
the composite operators at a quartic order in the constituent fields.

\bigskip

As we have seen, with the specific choice of composite operators discussed
here, of quartic order in the constituent spinon and holon fields~\footnote{%
The form of the quadratic parts (in the constituent fields) of these
operators was motivated by microscopic lattice system considerations~\cite%
{farakos,mav}.}, we have arrived at an $N=2$ supersymmetric extension which
coupled the two N=1 supersymmetries of the quadratic composites of \cite%
{compos}. The coupling was done in a way consistent with a supersymmetric
Abelian Higgs model~\cite{edelstein}. The emergence of a N=2 supersymmetry
algebra at a composite level is consistent with the elevation of N=1
supersymmetry to N=2 in the constituent theory~\cite{mav}, due to the
existence of topological currents in 2+1 dimensions.

It should be remarked at this stage that, from a physical point of view, one
starts from a microscopic lattice system, an appropriately extended $t-j$
model~\cite{mav} with nodes in its fermi surface. Upon assuming spin-charge
separation one arrives at a continuum low-energy theory of \textit{nodal}
spinons and holon excitations, which has the form of a relativistic $CP^1$ $%
\sigma$-model (magnon-spinons) coupled to Dirac-like fermions (holon degrees
of freedom). At supersymmetric points in the microscopic model phase space~%
\cite{mav} one recovers a supersymmetric theory between the spinon and holon
constituents without any dynamical gauge fields. This is the constituent
theory. The non-dynamical gauge fields simply express contact interactions
between spinons and holons.

Although at first sight the model appears to have only a $N=1$
supersymmetry, it actually has a hidden $N=2$ supersymmetry due to its low
dimensionality (2+1 dimensions), for reasons stated above. At the composite
operator level, obtained after integrating out the non-dynamical gauge
fields, one generates \textit{dynamical gauge fields}, made out of
appropriate combinations of spinon and holons. It is interesting to notice
that the choice of composite operators made in this paper seem to
necessitate an Abelian nature of the gauge field involved in the N=2
supersymmetric multiplet. We were unable to find a choice of composite
operators that generate the full non-Abelian SU(2) supersymmetric model of %
\cite{affleck}.

In view of this Abelian nature of the gauge group, an interesting question
arises as to whether the composite Abelian gauge field, which emerges from
this construction, is compact or not. The two cases lead to very different
non-perturbative dynamics~\cite{strassler}, as exemplified by the
confinement properties of the relevant three-dimensional gauge theory. Such
matters will be investigated in more detail in future works.

We believe that this work, together with our earlier works on the subject~%
\cite{mav,compos}, opens up a way for a formal discussion of exact
non-perturbative results for the phase-space dynamics of strongly correlated
electron systems, even if, in realistic situations, the latter lie away from
such supersymmetric points. If one understands analytically some aspects of
the non-perturbative dynamics at supersymmetric points, then one might hope
of using such knowledge to draw conclusions for the theory away from these
points, where the supersymmetry is explicitly broken. We intend to continue
working along these lines with a view to applying the results to realistic
condensed-matter situations of relevance to high-temperature
superconductivity, and more generally to antiferromagnetic systems.

\section*{Acknowledgements}

This work is supported by the Leverhulme Trust (U.K.).

\end{document}